\documentclass[preprint]{aastex}

\begin{document}

\title{AN UPPER BOUND TO THE SPACE DENSITY OF INTERSTELLAR  COMETS}

\author{M. Jura} 
\affil{Department of Physics and Astronomy, University of California, Los Angeles CA 90095}

\begin{abstract}
Two well-studied white dwarfs with helium-dominated atmospheres (DBs) each possess less hydrogen than  carried by a single average-mass comet.  Plausibly, the  wind rates from these stars are low enough that  most accreted hydrogen remains with the star.  If so, and presuming their nominal effective temperatures, then these DBs have  been
minimally impacted by interstellar comets during their 50 Myr cooling age; 
interstellar iceballs with radii between 10 m and 2 km   contain less than 1\% of all interstellar oxygen.  This analysis suggests that most stars do not produce comets at the rate predicted by ``optimistic"
scenarios for the formation of the Oort cloud. 
  \end{abstract}

\section{INTRODUCTION}

There are two reasons to imagine that interstellar comets could be widespread.  First, they could be an important as-yet undetected reservoir of interstellar oxygen.  Second, models for
the formation of the solar system's Oort cloud predict that more comets are ejected into the interstellar medium than
remain gravitationally bound to the Sun.  If our solar system is at all typical, then the birth of stars and planets might commonly lead to the creation of many interstellar comets.

Despite decades of effort, analysis of the rate of appearance of new comets in the solar system has not  provided a strong constraint on the space
density of interstellar comets.  Substantial progress may require another approach.   In this spirit,  \citet{Shull1995}  argued that one possible explanation for the ``low" observed frequency of gamma ray bursts arising from the Milky Way is that there are few interstellar comets impacting upon neutron stars.  However,  there are enough uncertainties in this approach, that their result that fewer than 4\% of stars have Oort clouds similar to the Sun's is not completely firm.  While also employing degenerate stars as a tool, here,  we proceed somewhat differently.  We argue that the small amounts of hydrogen accumulated by warm DB white dwarfs significantly limits the rate
of iceball impact  with these stars and therefore the  space density of interstellar comets.  Our method requires combining an
understanding of the interstellar medium with the physics of external pollution of white dwarfs.  We therefore subdivide this Introduction into subsections
to describe the different threads of our approach.

\subsection{Interstellar Oxygen}

Including both atoms  in the gas phase and those estimated to be contained within dust grains, the  abundance of oxygen within diffuse clouds, $n$(O)/$n$(H), is within 15\% of 5.1 ${\times}$ 10$^{-4}$ \citep{Cartledge2004}.  While this value  is in agreement with recent determinations of the solar photospheric oxygen abundance 
of $n$(O)/$n$(H) = 4.6 ${\times}$ 10$^{-4}$ \citep{Asplund2004} or  $n$(O)/$n$(H) = 5.8 ${\times}$10$^{-4}$ \citep{Caffau2008}, the true solar abundance may be as high as $n$(O)/$n$(H) = 7.2 ${\times}$ 10$^{-4}$ \citep{Delahaye2006}.  Even with the problematical assumption that the interstellar and solar oxygen abundances are the same,  the interstellar oxygen abundance is  ${\sim}$50\% uncertain.   The possibility of a missing reservoir of interstellar oxygen  has been considered for decades \citep{Greenberg1974}, and here we re-assess the hypothesis that appreciable oxygen is carried within interstellar comets.
Furthermore, although not directly related to the question of the total abundance of interstellar oxygen,   the observed decrease of  $n$(O)/$n$(H) with column density \citep{Cartledge2004,Jenkins2009} is not well understood and presents a ``crisis" in our understanding of the behavior and distribution  of interstellar oxygen \citep{Whittet2010}. 

With an
 interstellar density of hydrogen nuclei near  the Sun of 1.15 cm$^{-3}$ \citep{Bohlin1978} and with $n$(O)/$n$(H) = 5.1 ${\times}$ 10$^{-4}$,  the mass density of interstellar oxygen is 1.6 ${\times}$ 10$^{-26}$ g cm$^{-3}$.  Next, we  compare this estimate
 with observational bounds and theoretical expectations for the space density interstellar comets and the amount of oxygen that they contain.
\subsection {Interstellar Comets?}
 
 To date, all observed solar system comets appear
 to have always been gravitationally bound to the Sun. Consequently
 \citet{Whipple1975} and \citet{Sekanina1976} placed  upper limits of 2 ${\times}$ 10$^{-26}$ g cm$^{-3}$ and 4 ${\times}$ 10$^{-26}$ g cm$^{-3}$, respectively, to the mass density of interstellar comets. Using the same approach but with updated parameters, 
 \citet{Francis2005} placed an upper bound to the space density of interstellar comets of  4.5 ${\times}$ 10$^{-4}$ AU$^{-3}$.   With an average mass per comet as much as 1.2 ${\times}$ 10$^{18}$ g \citep{Francis2005}, and assuming about 50\% of a comet's mass  is oxygen  \citep{Greenberg1998}, then  the  upper bound to the  mass density of oxygen within interstellar comets
 is 8 ${\times}$ 10$^{-26}$ g cm$^{-3}$ -- larger than the amount of  gas-phase interstellar oxygen \citep{Stern1990b}.
 
 Conventional dynamic models for the formation of the Oort cloud from objects in the inner solar system \citep{Duncan1987} predict that there could be as many as 50  comets ejected into the interstellar medium for every   comet that remains bound to the Sun.  Unconventional models for the Oort cloud \citep{Levison2010}  also must require significant numbers of interstellar comets.  
 An ``optimistic" model for the space density of interstellar comets presumes that all stars form massive comet clouds similar to the solar system's Oort cloud and that most of these comets are lost into the interstellar medium.  With this picture, \citet{Stern1990a} predicted that the space density of interstellar comets could be as high as 2 ${\times}$ 10$^{-43}$ cm$^{-3}$ (or 7 ${\times}$ 10$^{-4}$ AU$^{-3}$), comparable to the observational upper limit of \citet{Francis2005}.  Consequently,  ``optimistic" models predict that  occasional interstellar comets can be observed within the solar system \citep{Mcglynn1989,Sen1993}.   However, since the mass of the Oort cloud, the fraction of comets that are ejected into the interstellar medium and the fraction of stars that create analogs to the Oort cloud are each uncertain by a factor of ${\sim}$10 \citep{Stern1990a,Shull1995,Weissman1996,Dones2004}, the ``optimistic"
 scenario may vastly overestimate the true space density of interstellar comets.     
 
 \subsection{A New Approach}
 The basic idea of this paper is to use white dwarfs with helium dominated atmospheres (DBs) to limit the space density of
 interstellar comets.
The large majority of  white dwarfs  have atmospheres which are hydrogen dominated and  these stars are classified
as DAs \citep{Liebert1986,Sion1988}.  The standard scenario is that gravitational settling of heavy elements is very effective, and therefore, with sufficient time, essentially all of the  hydrogen is located in the star's outer envelope and dominates the composition of the photosphere \citep{Hansen2003}.  A minority of white dwarfs have so little hydrogen that they have helium dominated atmospheres even though all their hydrogen resides in the star's outer envelope. If lines from neutral helium are detected, they are classified as DBs.  If too cool for neutral helium to be detectable yet they display heavy elements in their spectra, they are  classified as DZs.

Two recent surveys show that white dwarfs accumulate hydrogen from an external source because there is a a correlation between the average  hydrogen  mass and a DB's or DZ's cooling age \citep{Dufour2007,Voss2007}.   The typical rate accretion rate is near 6 ${\times}$ 10$^{5}$ g s$^{-1}$ \citep{Jura2010}, but there are large variations.
  In this paper, we focus on the ${\sim}$50\% of DBs with only upper bounds to the mass of accumulated hydrogen \citep{Voss2007}.

The mass of the convective zone where the accreted material resides is a strong function of
the star's effective temperature.  Because  DBs with $T_{*}$ $<$ 20,000 K have convective zones that are 10,000 times more
massive than  DBs with $T_{*}$ $>$ 22,000 K \citep{Koester2009}, these two temperature classes may display external pollution quite
differently.

There is relatively little 
hydrogen in DBs warmer than 22,000 K.   A conservative upper bound from optical and ultraviolet observations of most ``warm"
DBs  is that $n$(H)/$n$(He) $<$  10$^{-4}$ \citep{Beauchamp1999,Voss2007}.  Since the mass of the outer convective zones of these warm DBs  is at most 5 ${\times}$ 10$^{21}$ g \citep{Koester2009}, these stars have accumulated less than  ${\sim}$10$^{17}$ g of hydrogen.  As described in Section 3, two particularly well studied DBs, GD 190 and BPM 17088, have accumulated less than 4 ${\times}$ 10$^{15}$ g of hydrogen, roughly the amount of this element carried in a single average-size comet.  At the ``typical" accretion rate of 6 ${\times}$ 10$^{5}$ g s$^{-1}$ experienced by the cool DBs, these two stars 
would accumulate the upper bound to their hydrogen in less than 10$^{3}$ yr, much less time than their cooling ages of  50 Myr.  Either the warm DBs do not accrete at the same rate as many cooler DBs, or they lose almost all their acquired hydrogen in a wind.

In Section 2 we elaborate in detail upon the  ``naive" idea that
 warm DBs have not accumulated much hydrogen simply because they have not accreted  from ice-rich asteroids, comets or gas from the interstellar medium -- the current best candidates as the sources for hydrogen in  cooler DBs. While this argument hardly seems remarkable, there are interesting consequences.  Using  the toy model
 for a DB's hydrogen budget  developed in Section 2,   
  in Section 3 we
constrain the mass carried within interstellar comets.  In Section 4, we outline observational possibilities
 while in Section 5 we  present our conclusions.

\section{TOY MODEL FOR A  WARM DB'S HYDROGEN BUDGET}

\subsection{Model Description}
We  compute the amount of hydrogen in ``warm" DBs for  a regime  when essentially all of its hydrogen of mass, $M(H)$ has risen to the surface layers and is well mixed in the outer convective zone of mass $M_{cvz}$.  There may be a very thin
non-convective photosphere lying above the convective zone, but  we assume that the photosphere and
the convective zone have the same composition \citep{Dupuis1992,Su2010}. 
 If hydrogen is
accreted with rate ${\dot M}_{acc}(H)$ and the total mass loss rate is ${\dot M}_{wind}$, then:
\begin{equation}
{\dot M(H)}\;=\;{\dot M_{acc}(H)}\,-\,\frac{M(H)}{M_{cvz}}\,{\dot M}_{wind}
\end{equation}
 Equation (1) is mathematically equivalent
to the expression governing the abundance of heavy atoms in the outer layer of a white dwarf where the atoms are ``lost" because of downward settling (Koester 2009) instead of our case where
hydrogen atoms are lost in an outward flow in a  wind.

For our purposes here, there is a crucial difference between accretion of gas from the interstellar medium and accretion from large solid objects such as asteroids.
There is no reason to think that a quasi-spherical stellar wind strongly inhibits the accretion from a flat disk produced by tidal disruption of a  parent body \citep{Jura2003}.  In contrast, such a  wind can physically block interstellar gas accretion.  In particular, while winds of 10$^{6}$ g s$^{-1}$ can prevent interstellar
accretion at the typical  rate, once accreted hydrogen is diluted into a helium-dominated envelope, much higher wind outfows are required to
inhibit the build-up of observable amounts of hydrogen.  

It is useful to introduce the characteristic time for the wind to deplete the mass in the convective zone, $t_{depl}$:
\begin{equation}
t_{depl}\;=\;\frac{M_{cvz}}{{\dot M}_{wind}}
\end{equation}
If $M_{cvz}$, ${\dot M}_{wind}$ and ${\dot M}_{acc}$ are independent of time, the  solution to Equation (1) is:
\begin{equation}
M(H)[t]\;=\;M(H)[0]\;e^{-t/t_{depl}}\;+\;{\dot M}_{acc}(H)\,t_{depl}\,\left(1\,-\,e^{-t/t_{depl}}\right)
\end{equation}
Although Equation (3) provides a guide to our results,  it is inexact because $M_{cvz}$ is not constant during the star's cooling.  
In the numerical solutions to Equation (1) presented below, we fit the detailed calculations for $M_{cvz}$  in a 0.6 M$_{\odot}$ DB without hydrogen  by \citet{Koester2009} and \citet{Gautschy2002} in the
temperature interval 22,000 K $<$ $T_{*}$ $<$ 28,000 K with the expression:
\begin{equation}
M_{cvz}\;{\approx}\;1.5\,{\times}\,10^{21}\,\left(\frac{25,000}{T_{*}}\right)^{10}\;\;(\rm{g})
\end{equation}
Also, we relate the star's effective temperature to its cooling age by fitting the calculations of Bergeron\footnote{see http://www.astro.umontreal.ca/${\sim}$bergeron/CoolingModels/Tables/Table$\_$DB} for a 0.6 $M_{\odot}$ DB white dwarf with:
\begin{equation}
 t\; {\approx} \;49.7\; \left(\frac{22,000}{T_{*}}\right)^{5}\;\;  (\rm{Myr})
\end{equation}
Because $M_{cvz}$ increases as the star ages, then, as can be seen from Equation (1), other factors being equal, the loss rate of hydrogen is diminished.
Consequently, instead of reaching an asymptotic value as predicted by Equation (3), as illustrated in Figures 1-3, $M(H)$ continues to increase with time.
\subsection{Boundary Condition}

We set $t$ = 0 for  $T_{*}$ = 28,000 K, since at this moment in a DB's cooling, $M_{cvz}$ becomes large enough that
Equation (1) can be sensibly employed.  The value of 
$M(H)[0]$ is sensitive to the white dwarf's previous history \citep{Unglaub2000}.  However, since the time for a 0.6 M$_{\odot}$ DB to cool from
$T_{*}$ = 28,000 K to $T_{*}$ = 22,000 K is about 35 Myr, then in most cases of interest, $t$ $>$ $t_{depl}$ since we focus on wind loss
rates greater than 10$^{6}$ g s$^{-1}$.  
If so, then the exact value of $M(H)[0]$ is unimportant.  In all cases, we adopt 
$M(H)[0]$ = 1 ${\times}$ 10$^{15}$ g,  a conservative upper bound for the mass of hydrogen in  EC  20058-5234 \citep{Petitclerc2005}, a well studied DB with $T_{*}$ = 28,000 K \citep{Sullivan2008}. 

 \subsection{White Dwarf Winds?}
 
 By far, the largest uncertainty  in  Equation (1) is the stellar wind rate from white dwarfs.   \citet{Unglaub2008} has computed that radiatively-driven winds
from white dwarfs with $T_{*}$ $<$ 30,000 K do not occur.  Quite possibly, therefore, we should adopt ${\dot M}_{wind}$ = 0.  In contrast, in order to explain the presence of small amounts of dredged-up carbon in warm DBs, \citet{Fontaine2005} have proposed that ${\dot M}_{wind}$ is as high as ${\sim}$10$^{13}$ g s$^{-1}$ for $T_{*}$ $>$ 20,000 K.  However, the physical explanation for this putative wind is not provided.  One possibility might be that the white dwarfs have hot coronae which could drive a wind \citep{Lamers1999}.  In fact, white dwarfs do have convective envelopes and there could be some transfer of the convective energy
flux into heating the outer layers of the gas.   

We use two lines of evidence  to bound the wind rates from warm white dwarfs.  First, these stars  are not X-ray sources \citep{Odwyer2003}.   Assume a wind outflow speed  near the white dwarf's characteristic escape velocity, $V_{esc}$, of  4000 km s$^{-1}$ \citep{Lamers1999} and consider the kinetic energy flow carried by 
 the wind, $1/2\,{\dot M}_{wind}\,V_{esc}^{2}$.  Much of  this energy  is deposited into the surrounding interstellar medium to create a hot bubble.  Using the result that $L_{X}$ $<$ 4 ${\times}$ 10$^{27}$ erg s$^{-1}$ \citep{Musielak2003} for GD 358, a DB with $T_{*}$ = 27,000 \citep{Provencal2009}, it seems  that ${\dot M}_{wind}$ $<$  ${\sim}$ 5 ${\times}$ 10$^{10}$ g s$^{-1}$.  (This limit could be relaxed if the X-rays are deposited in a region greater than the 5{\farcs}1 circle on the sky employed by \citet{Musielak2003} in their data analysis.  However,
 as can be seen from their Figure 2, there is no extended X-ray emission.)

 Second, some white dwarfs accrete and therefore are not losing a large amount of mass in a wind.   
 Consider
 SDSS 1228+1040 with $T_{*}$ = 22,000 K,  the hottest white dwarf known to be heavily polluted and possess a dust disk (Gaensicke et al. 2006, Brinkworth et al. 2009, Melis et al. 2010).  Because this DA does not have a convective envelope \citep{Koester2009},   Equation (1) does not apply.  Instead, to accumulate heavy elements,  it is necessary that ${\dot M}_{wind}$ $<$ ${\dot M}_{acc}$.
 From the formalism and parameters given by \citet{Koester2009}, the magnesium abundance
 reported by \citet{Gaensicke2007} and the presumption that ${\sim}$10\% of the mass of the material is magnesium (see, for example, \citet{Klein2010}), then the heavy atom accretion rate is about 2 ${\times}$ 10$^{9}$ g s$^{-1}$.  This value is within the range of 3 ${\times}$ 10$^{8}$ g s$^{-1}$ to ${\sim}$ 10$^{10}$ g s$^{-1}$ inferred for most white dwarfs with dust disks \citep{Kilic2006,Vonhippel2007,Farihi2009}. It seems likely that the wind rate from this star is less than 10$^{10}$ g s$^{-1}$. (We assume
 that this DA without a convective envelope has the same mass loss rate as a DB with a  low mass convective envelope.  Therefore, this
 argument is not completely certain.)
 
The source of the hydrogen in  DBs with $T_{*}$ $<$ 20,000 K Is uncertain.  If some of these stars accrete interstellar gas with the typical time-averaged hydrogen accretion rate given by  \citet{Voss2007} of ${\sim}$10$^{6}$ g s$^{-1}$, then ${\dot M}_{wind}$ likely is less
than this value.    Because warm DBs have
 much less massive convective envelopes than cool DBs, there is no reason to imagine that ${\dot M}_{wind}$ increases dramatically with effective temperature.  
We conclude  that  the wind rates from warm DBs likely are less than 10$^{10}$ g s$^{-1}$ and plausibly less
than 10$^{6}$ g s$^{-1}$.

Besides a wind, white dwarfs also may lose mass in explosive events as occurs in novae.  Because there is no evidence  that isolated white dwarfs undergo such events nor is there any theoretical reason to expect them,  we ignore this possibility.

\subsection{Interstellar Gas Accretion?}

Although considered for many years, the physics of accreting interstellar hydrogen gas  by white dwarfs is not well understood.  It seems, for example,  that accretion at the Bondi-Hoyle
rate is excluded \citep{Farihi2010}. DBs younger than 50 Myr  have  performed
less than one full oscillation through the Galactic Plane.  Therefore, these stars  may have only encountered hot, low density material and consequently have  accreted interstellar gas at a negligible rate during their entire cooling age.  Because  accretion of interstellar gas-phase hydrogen can only
strengthen the upper limits to asteroidal and cometary accretion discussed below, we  neglect this possibility.

 \subsection{Accretion of asteroids with ice?}
 
 In our calculations for asteroidal accretion, we assume that ${\dot M}_{acc}(H)$ = 6 ${\times}$ 10$^{5}$ g s$^{-1}$, a characteristic hydrogen accretion rate for
  DBs with $T_{*}$ $<$ 20,000 K although there is a large variation in this rate \citep{Voss2007}.
    We show in Figure 1 more detailed solutions to Equation (1).   In these models, we see that  if  ${\dot M}_{wind}$ $<$ 10$^{10}$ g s$^{-1}$, then DBs typically accumulate more hydrogen than observed.  This means that DBs warmer than 22,000 K either have ``strong" winds or they do not accrete from ice-rich asteroids.  

\subsection{Oort Cloud Analog Cometary Accretion?}

 \citet{Alcock1986} proposed that direct impact by comets can produce observable signatures in a white dwarf's photosphere.  Here, we estimate the
  hydrogen pollution of a DB white dwarf  by comets in an analog to the solar system's Oort cloud.  According to \citet{Francis2005}, 0.8  dynamically new comets AU$^{-1}$ yr$^{-1}$ arrive in the inner solar system.
  With a flat distribution of perihelia, the expected rate of collisions onto a star of radius 0.013 R$_{\odot}$ is 4.8 ${\times}$ 10$^{-5}$ yr$^{-1}$.  With an average comet mass  somewhere between 5.6 ${\times}$ 10$^{16}$ g and 1.2 ${\times}$ 10$^{18}$ g (Francis 2005) and the assumption that 5\% of the comet's mass is hydrogen \citep{Greenberg1998}, then the hydrogen accretion rate is expected to
 lie between 4 ${\times}$ 10$^{3}$ g s$^{-1}$ and 9 ${\times}$ 10$^{4}$ g s$^{-1}$.  To be conservative, we adopt the lower bound
 and show results from Equation (1) in Figure 2.  We see that if ${\dot M}_{wind}$ $<$ 10$^{8}$ g s$^{-1}$, then as long as $T_{*}$ $>$ 22,000 K, the star accumulates more  hydrogen than allowed for  the typical upper bound for DBs obtained from optical spectra \citep{Beauchamp1999}.  Therefore, if the winds from these white dwarfs are plausibly weak, the mass of their analogs to Oort cluds are smaller than the solar system's.  Either these stars 
 never had many primoridal comets or  they lost their comets when the star evolved on the Asymptotic Giant Branch \citep{Parriott1998}. 
 
  We assume that the comet's hydrogen is visible in the star's outer convective zone.  In fact, during its high-speed impact, the iceball might penetrate beneath the convective zone. Following
\citet{Alcock1986}, we assume that the material vaporized from the comet has  such a high entropy that it quickly rises to the surface of the star and therefore could be detected.  

\section{LIMITS ON INTERSTELLAR COMETS}

Having argued above that the well known sources of hydrogen accretion are not important in many  warm DBs, we now proceed to estimate the space density of interstellar comets.  

\subsection{Computed Accretion Rate} 
  We assume that the comets have hyperbolic orbits with  speed relative to
the white dwarf of $V_{wd}$.  If the comet impacts a white dwarf of radius $R_{*}$ and mass $M_{*}$, then it is 100\% accreted; we assume there is no accretion from a near-miss.  We treat the ambient interstellar iceballs as a continuous fluid with a hydrogen mass density, ${\rho}_{ball}(H)$.
Because the escape speed from the surface of the white dwarf is much greater than $V_{wd}$, then the total mass of accreted hydrogen, $M_{acc}(H)$ is (see, for example, \citet{Spitzer1978}):
\begin{equation}
M_{acc}(H)\;{\approx}\;2\,{\pi}\,G\,M_{*}\,R_{*}{\int} \frac{{\rho}_{ball}(H)[t]}{V_{wd}(t)}\,dt
\end{equation}
Below, we describe how
 ${\rho}_{ball}$ and $V_{wd}$ may vary with time.

Even before impact,  incoming comets may split  as they are heated and/or tidally disrupted.  However, the resulting fragmentation speeds of ${\sim}$0.1 km s$^{-1}$ \citep{Sekanina1982} are vastly smaller than the impact speeds that exceed 1000 km s$^{-1}$.  We therefore neglect the
small deviations from the initial hyperbolic orbits that arise.  In any case, the loss of some fragments that deviate away from a path that leads them into a direct
collision is balanced by the gain of other fragments from parent bodies on a near-miss trajectory that deviate into a collision-course.

For simplicity, we assume $V_{wd}$ is dominated by the white dwarf's space motion in the Local Standard of Rest and we ignore the motions of the comets.
Assuming that only the motion vertical to the plane varies significantly with time, we write:
\begin{equation}
V_{wd}\;=\;\left(U^{2}\,+\,V^{2}\,+\,W(t)^{2}\right)^{1/2}
\end{equation}
with $U$, $V$ and $W$ having their conventional meanings.  
We adopt the simplifying assumption that  relative to the Galactic Plane a star undergoes vertical simple harmonic motion \citep{Kuijken1989} 
\begin{equation}
W(t)\;=\;W_{0}\,\cos\,k^{1/2}\,t
\end{equation} 
With $k^{1/2}$ = 2.3 ${\times}$ 10$^{-15}$ s$^{-1}$ \citep{Siebert2003}, 
the period of a star's oscillation through the Galactic Plane is  86 Myr.

We assume that
interstellar iceballs have the same  spatial distribution as F-type stars and therefore  relative to the Galactic Plane:
\begin{equation}
{\rho}_{ball}(H)[z]\;=\;{\rho}_{ball}(H)[0]\,e^{-|z|/h}
\end{equation}
We adopt $h$ = 150 pc \citep{Gilmore2000}.

Our analysis only applies to iceballs in a limited size range. Small iceballs are significantly sublimated on their approach to the white dwarf, and the resulting gas atoms
may not accrete onto the star.  
Reformulating slightly Equation (14) in \citet{Jura2005}, the thickness, ${\Delta}R$, of a comet which is thermally sublimated as it approaches a star is:
\begin{equation}
{\Delta}R\;{\approx}\;\frac{L_{*}\,m_{H_{2}O}{\Delta}{\phi}_{subl}}{16{\pi}{\Delta}E[2\,G\,R_{*}\,M_{*}]^{1/2}}
\end{equation}
where $m_{H_{2}O}$ is the mass of a water molecule, ${\Delta}{\phi}_{subl}$ is the angular variation in its orbit traced by the incoming comet as seen by the host star where ice sublimation dominates the cooling of the white dwarf, ${\Delta}E$ is the energy per molecule  associated with sublimation (taken equal to 2 ${\times}$ 10$^{-12}$ erg, \citet{Sekanina2002}) and we assume an orbit that is nearly parabolic.  We take ${\Delta}{\phi}_{subl}$ = ${\pi}/2$ since  ice  sublimates significantly at distances much greater than $R_{*}$.  For illustrative purposes, we take $L_{*}$ = 0.05 L$_{\odot}$, $R_{*}$ = 0.013 R$_{\odot}$ and  $M_{*}$ = 0.6 M$_{\odot}$ as appropriate for the white dwarfs of interest.   We find that ${\Delta}R$ ${\approx}$ 240 cm.  Therefore, iceballs with
radii greater than ${\sim}$10 m deposit  most of their mass onto the white dwarf.  Our analysis only constrains iceballs with hydrogen masses less than the upper limits of accreted hydrogen
in our target white dwarfs.  This implies that we consider iceballs with less than 4 ${\times}$ 10$^{15}$ g of hydrogen and therefore a total mass less than 8 ${\times}$ 10$^{16}$ g and a radius less than 2 km.

Besides thermal sublimation, is there some other process which leads to a comet's disintegration and would 
  the  resulting gas and debris somehow be inhibited from accreting onto the star?  In considering this possibility, 
it should be recognized that the approach of a comet to a white dwarf is less disruptive than the approach of a comet to the Sun for two reasons.  First, the white dwarf's luminosity is
lower than the Sun's and thus a comet nearing a white dwarf is heated less and the resulting thermal stresses are smaller. Second, the typical white dwarf mass of 0.6 M$_{\odot}$ is less than the Sun's mass and therefore the 
gravitational tidal forces from a white dwarf are smaller.  As a result, while comets may split as they approach their target white dwarf, analogous to the approach
of comets to the Sun, we do not expect them to fully disintegrate -- if they ever do -- until they are closer than ${\sim}$ 1 $R_{\odot}$.  At this separation, the comet's incoming speed is nearly 500 km s$^{-1}$ and it survives less than 1000 s before
impact.  Perhaps as the comet nears within 1 R$_{\odot}$ of the white dwarf, there may be extensive  explosions.  Generalizing from the observations of comet 17P/Holmes which displayed remarkable transient brightening \citep{Reach2010}, the typical outflow speed of dust fragments and gas debris in such events might be 0.1 km s$^{-1}$.
At this speed, the ejected material  travels ${\sim}$100 km in the comet's frame of reference before impacting the white dwarf.  Since this
transverse distance is much smaller than the radius of the white dwarf of nearly 10$^{4}$ km, we expect that cometary material that initially is on a trajectory aimed at the star does in
fact impact, regardless of any explosions that occur during the last 1000 s.  This argument applies even to small dust grains.     The luminosity of a white dwarf is so low that even tiny solid particles  cannot be substantially deflected by
radiation pressure (see Section 5.6 of \citet{Plavchan2005}).  In conclusion,  even if a comet largely disintegrates during its final
approach to a white dwarf, the bulk of the material is accreted.

\subsection{Application to GD 190 and BPM 17088}

To date, the most sensitive upper limits to  hydrogen in DBs are by  \citet{Petitclerc2005} who report ultraviolet spectroscopy for three stars.    One of the stars, EC 20058-5234 with  $T_{*}$ = 28,000 K, is relatively young and just beginning to have 
a well developed convective zone \citep{Gautschy2002}.   Our toy model described in Section 2 may not apply.  Therefore, we only  quantitatively discuss results for GD 190 and BPM 17088.  
  The parameters that we adopt for the two targets and relevant references are given
in Table 1.   If their effective temperatures are less than 22,000 K, then our derived upper limits to the space density of interstellar comets are weakened.

From Equation (8), the maximum distance that a star travels from the Galactic Plane is $W_{0}/k^{1/2}$.  If $W_{0}$ equals the value of $W$ given in Table 1, then GD 190 and BPM 17088 remain within 150 pc, the estimated scale height around the Galactic Plane of the interstellar iceballs.   Using the parameters in Table 1,   and making the simplifying assumption that the interstellar density and speed of the white dwarf  are constant since  $|W(t)|$ $<<$ $V_{wd}$  and $|z|$ $<$ $h$, then from Equation (6),
 the upper limits to  ${\rho}_{ball}(H)$ experienced by  GD 190 and BPM 17088 are 1.5 ${\times}$ $10^{-29}$ g cm$^{-3}$ and 1.0 ${\times}$ 10$^{-29}$ g cm$^{-3}$, respectively.   We assume that interstellar comets are mainly ice and silicates and that 
${\rho}(O)$ ${\approx}$ 10 ${\rho}(H)$ (Greenberg 1998).  Consequently
the upper bounds to ${\rho}_{ball}(O)$ in the Galactic midplane are 1.5 ${\times}$ 10$^{-28}$ g cm$^{-3}$ and 1.0 ${\times}$ 10$^{-28}$ g cm$^{-3}$ for GD 190 and BPM 17088, respectively.  Therefore, we find that the density of oxgyen contained within large iceballs is less than about 1\% of the total interstellar oxygen density, and that  ``optimistic"
models for the space density of interstellar comets (Section 1.2) with radii between 10 m and 2 km are excluded.   

Above, we assumed that any wind from the white dwarf is completely negligible.  We now consider the implications of including a wind and show in Figure 3  results for the amount
of  hydrogen accumulated by a DB as a function of time for a case where the average accretion rate is 10 g s$^{-1}$, a factor of four above the values given in Table 1.  We see that if ${\dot M}_{wind}$ is less than 10$^{6}$ s$^{-1}$, more hydrogen is accumulated than allowed by the observed upper bound of
4 ${\times}$ 10$^{15}$ g.  Therefore, if warm white dwarf winds are plausibly weak, interstellar iceballs are rare.

\section{OBSERVATIONAL POSSIBILITIES}

We have suggested that DBs warmer than 22,000 K are not polluted by ice-rich asteroids, comets or interstellar gas.    This scenario can be tested by examining
 atmospheric compositions.  Using
ground-based observations, \citet{Zuckerman2010} found that over 30\% of DBs between 13,500 K and 19,500 K display heavy atoms.  To date, the
most extensive similar study of DBs with $T_{*}$ $>$ 22,000 K is by \citet{Voss2007}.   In this sample of 11 stars, only WD 2354+1659 was identified as having heavy atoms in its photosphere.  One possibility is that  this star has accreted ice-poor asteroids as could be inferred with a detailed abundance analysis similar to that performed by \citet{Klein2010} for GD 40.  
Also, however, the temperature of this star is uncertain.  Depending upon the gravity and hydrogen content, $T_{*}$ could be as high as 24,800 K \citep{Voss2007} or as low as 19,000 K \citep{Koester2005}. While the frequency of externally-polluted ``warm" DBs appears to be low compared to the cooler DBs, further  measurements should be performed since the spectra obtained by \citet{Voss2007} are not as sensitive to heavy atoms  as those obtained by \citet{Zuckerman2010}.

Ultraviolet spectroscopy is a sensitive tool to measure heavy elements in a white dwarf's atmosphere \citep{Desharnais2008}.  Data from the {\it Far Ultraviolet Spectroscopic Explorer} and  the {\it Hubble Space Telescope} have been obtained for
5 DBs warmer than 22,000 K: GD 190, BPM 17088, EC 20058-5234 \citep{Petitclerc2005}, PG 0112+104 \citep{Provencal2000} and GD 358 \citep{Provencal1996}.  Although these stars possess
atmospheric carbon, it is almost certain that this heavy element is intrinsic to the star and not accreted (Fontaine \& Brassard 2005, Desharnais et al. 2008).  With the models of \citet{Koester2009}, the upper limits to the iron abundance found
by \citet{Petitclerc2005} and the assumption that 1/3 of the accreted mass is iron as seems appropriate for extrasolar asteroids \citep{Klein2010,Zuckerman2007, Zuckerman2010,Dufour2010,Vennes2010}, then in GD 190, for example,
the upper limit to the heavy atom accretion rate is ${\sim}$1 ${\times}$ 10$^{6}$ g s$^{-1}$, a  value lower than any reported by \citet{Zuckerman2010} for polluted cooler DBs. Therefore ultraviolet results are  more sensitive than optical studies for identifying pollution.  If each ultraviolet targeted DB has at least a 30\% probability
of experiencing pollution, as found by \citet{Zuckerman2010} for cool DBs, then the total probability that all 5 atars are uncontaminated is less than (0.7)$^{5}$ or 16\%.  Future observations of warm DBs may further test the prediction that these stars do not accrete
ice-rich asteroids.

The warmest DA known to have a dust disk is SDSS 1228+1040 with $T_{*}$ = 22,600 K \citep{Gaensicke2006,Brinkworth2009,Melis2010}. Discovery of additional warm DAs with dust and then measuring their pollution may
help extend the argument described in Section 2 to constrain the mass loss rates from such stars.  

In our toy model described in Section 2, we have treated hydrogen accretion as a minor perturbation in a DB atmosphere.
In fact, it may take only ${\sim}$ 10$^{18}$ g of hydrogen to prevent the conversion of a  DA into a DB for stars warmer than 22,000 K by dredge-up of interior helium \citep{Macdonald1991,Su2010}.  In contrast, for cooler stars with much more massive helium convective zones, considerably  more hydrogen may be required to prevent 
the conversion from DAs to DBs \citep{Macdonald1991,Tremblay2008}. The fraction of all white dwarfs that are DBs with $T_{*}$ between 25,000 K and 40,000 K
is a factor of 2.5 less than the same fraction of DBs with $T_{*}$ between 16,000 K and 22,000 K \citep{Eisenstein2006}.  We speculate
that the systems with ice-rich asteroids accrete enough hydrogen to prevent a white dwarf from becoming a DB until the temperature
falls appreciably below 22,000 K.  

Our supposition that the Oort cloud is unusually massive is consistent with the previous suggestion by \citet{Shull1995}, and the picture  that the solar system is somewhat special.  The orbital eccentricities of our planets are unusually low \citep{Marcy2000} and extrasolar analogs to the Late Heavy Bombardment  appear to be rare \citep{Booth2009}.

\section{CONCLUSIONS}

We identify a subclass of warm DBs where accretion of asteroids, analogs to Oort clouds comets and interstellar gas all appear to
be negligible.  
We make the plausible  argument that the wind mass loss rate these stars is less than 10$^{6}$ g s$^{-1}$.  
If these stars have their nominal effective temperatures, it then follows that because they have accumulated only a small mass  of hydrogen, interstellar comets with radii between 10 m and 2 km a contain less than 1\% of all interstellar oxygen.
``Optimistic" scenarios for the formation of the solar system's Oort cloud may not apply to most other stars.

This work has been partly supported by the National Science Foundation.  I thank B. Klein, M. Kilic, D. Koester, S. Xu and B. Zuckerman for helpful
comments.

\newpage
\begin{center}
Table 1 -- Target White Dwarfs

\begin{tabular}{llll}
\hline
\hline
 & GD 190 & BPM 17088 &Reference(s) \\
 \hline
 $T_{*}$ (K)& 22,000 & 22,600 & $^{a}$ \\
 $\log$ $g$ (cgs) & 7.97 & 8.07 & $^{a,b}$ \\
 $M_{*}$ (M$_{\odot}$) & 0.59 & 0.64 & $^{a,b}$ \\
 $\log$ $n$(H)/$n$(He) & ${\leq}$ -5.5 & ${\leq}$ -5.5 & $^{a}$ \\
 $\log M_{CVZ}$ (g) &   21.73 & 21.61 & $^{c}$ \\
 $\log M_{acc}(H)$ (g) & ${\leq}$15.63 & ${\leq}$ 15.51 \\
 $\log t_{cool}$ (Myr) & 7.70 & 7.64 & $^{d}$ \\
 $\log R_{*}$ (cm) & 8.96 & 8.93&  \\
${\dot M}_{acc}(H)$ (g s$^{-1})$& ${\leq}$2.7 & ${\leq}$2.4 \\
 $U$ (km s$^{-1}$) &-19  & 14 & $^{e}$\\
 $V$ (km s$^{-1}$) & -12 & -10 & $^{e}$  \\
 $W$ (km s$^{-1}$) & -9 & 8 & $^{e}$\\
 $V_{wd}$ (km s$^{-1}$) & 24 & 19 & \\
  \hline
 \end{tabular}
 \end{center}
 $^{a}$\citet{Petitclerc2005}, $^{b}$\citet{Voss2007}, $^{c}$Equation (4) and ignoring the mass variation of this quantity,
 $^{d}$Equation (5) and ignoring the mass variation in this quantity,
$^{e}$\citet{Sion1988}

 \newpage
 \begin{figure}
 \plotone{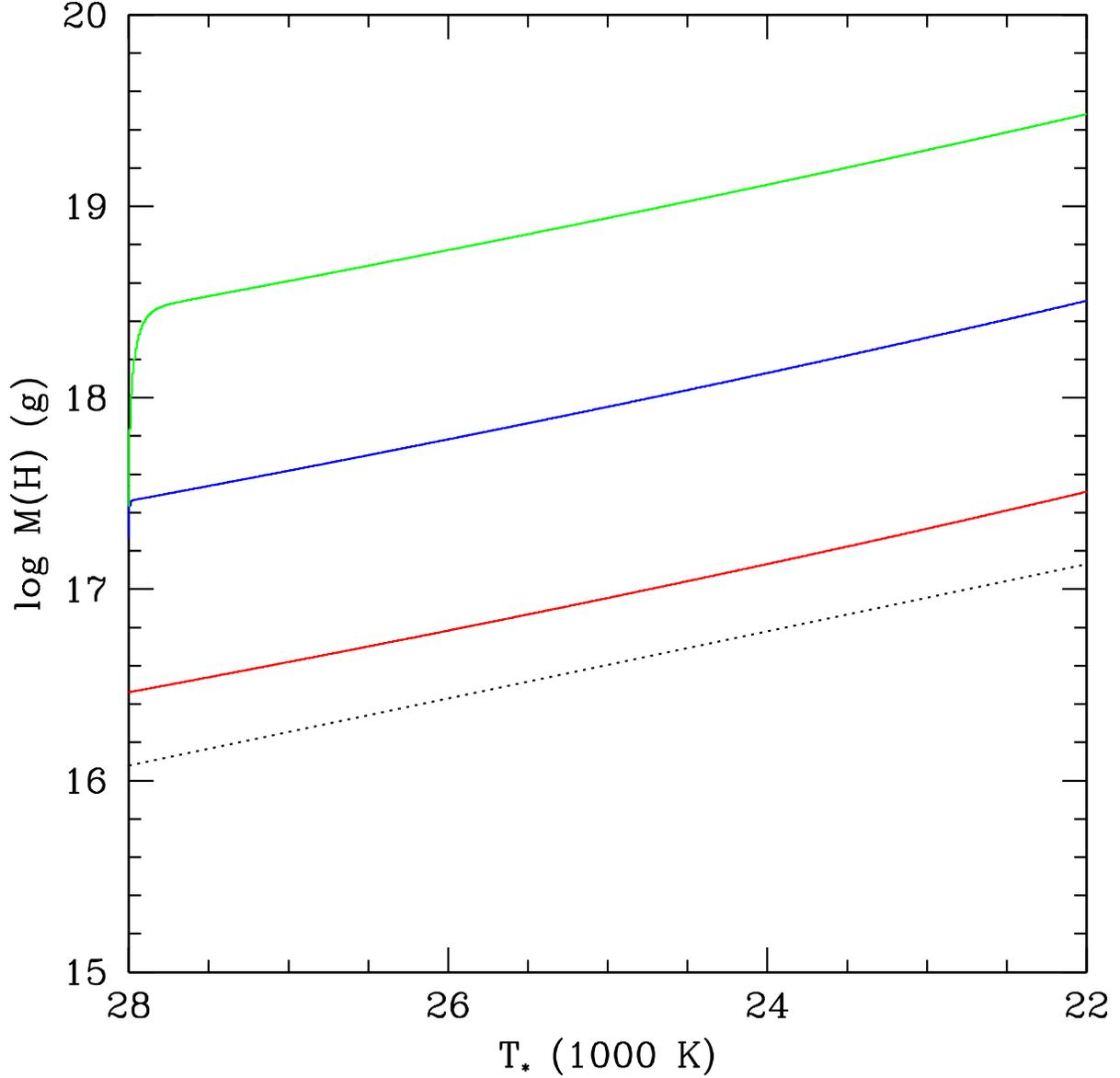}
 \caption{Predicted accumulated hydrogen mass from asteroidal accretion as a function of effective temperature for different wind mass loss rates.  We use $M(H)[0]$ = 1.0 ${\times}$ 10$^{15}$ g, ${\dot M}_{acc}(H)$ = 6 ${\times}$ 10$^{5}$ g s$^{-1}$ and values of
 ${\dot M}_{wind}$ of 10$^{10}$, 10$^{9}$ and 10$^{8}$ g s$^{-1}$ shown by the red, blue and green lines, respectively.  The dotted line represents the  upper limit to hydrogen in a DB with the assumptions that $n$(H)/$n$(He) ${\leq}$ 10$^{-4}$ and $M_{cvz}$ given by Equation (4).  We see that in the likely case that  ${\dot M}_{wind}$ $<$10$^{10}$ g s$^{-1}$, the assumption that the star accretes ice-rich asteroids
over-predicts the amount of accumulated hydrogen}
 \end{figure}
 \begin{figure}
 \plotone{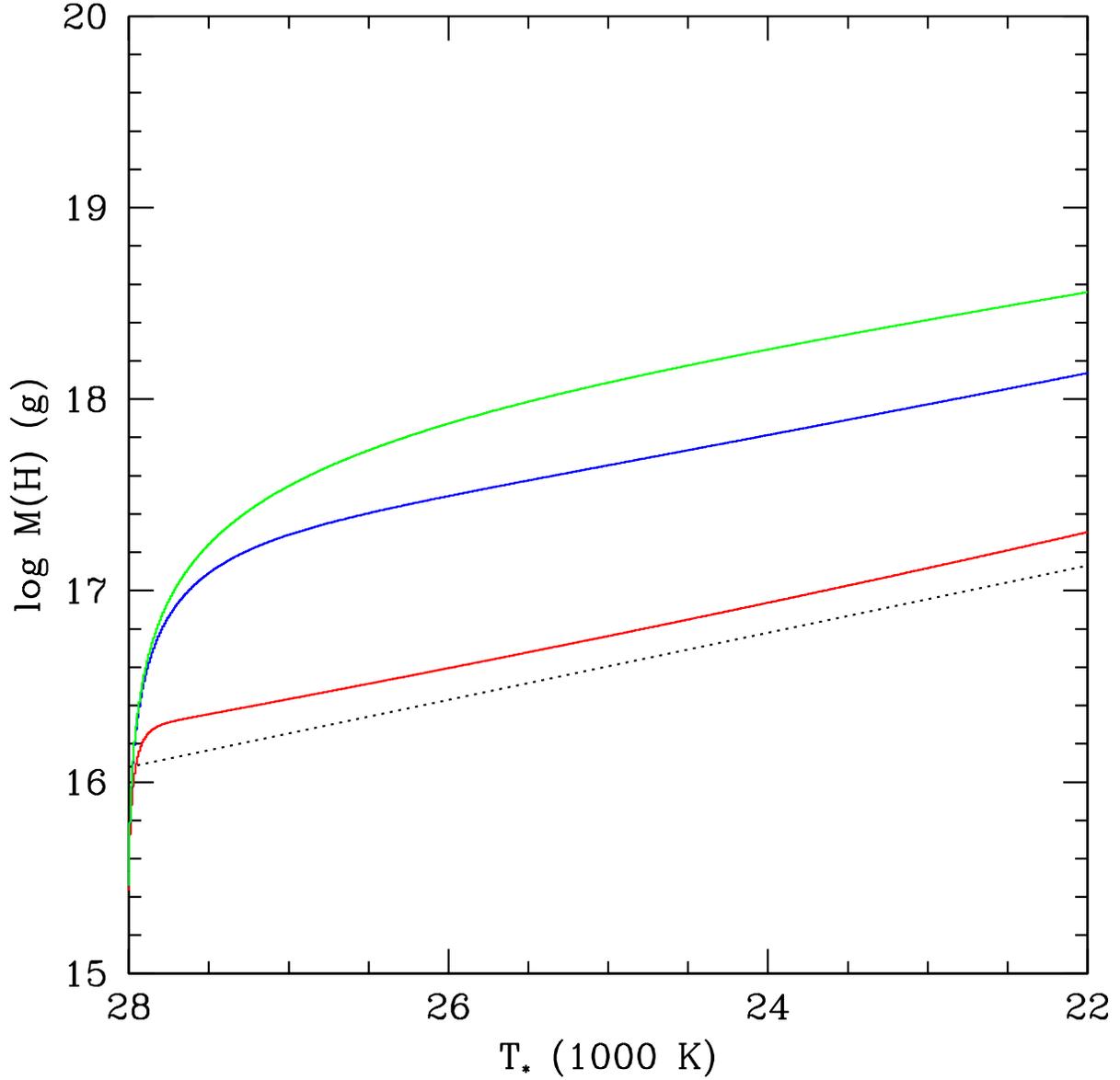}
 \caption{Similar to Figure 1 except for accretion from an analog to the Oort cloud with ${\dot M}_{acc}(H)$ = 4 ${\times}$ 10$^{3}$ g s$^{-1}$ and values of ${\dot M}_{wind}$ of 10$^{8}$, 10$^{7}$ and 10$^{6}$ g s$^{-1}$ shown by the red, blue and green curves, respectively.  We see that
 in the plausible case that ${\dot M}_{wind}$ $<$ 10$^{8}$ g s$^{-1}$, the assumption that the gas accretes comets from a conservative extrapolation of our own Oort cloud  over-predicts the amount of accumulated hydrogen.}
 \end{figure}
 \begin{figure}
 \plotone{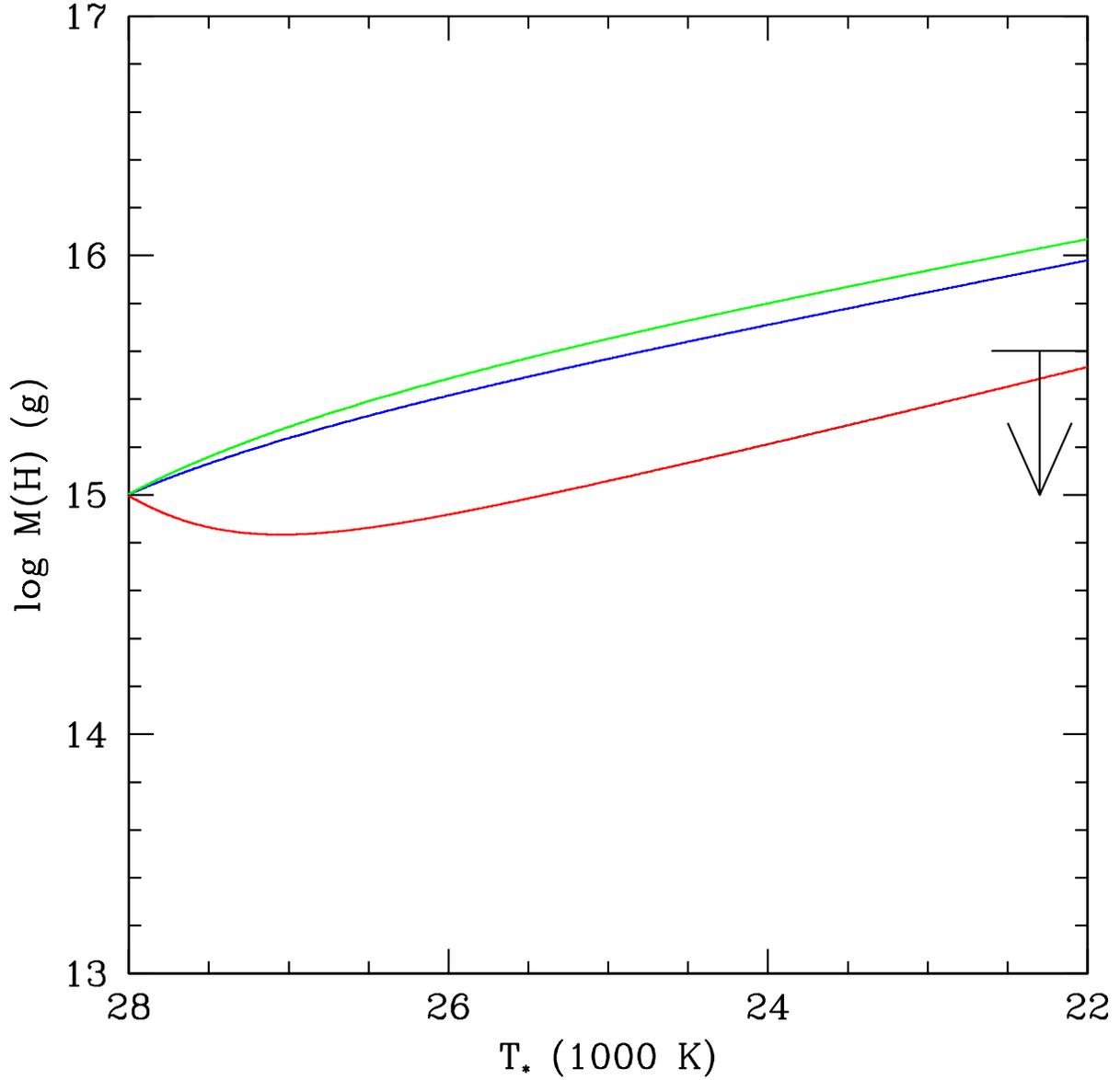}
 \caption{Similar to Figure 1 except for an accretion rate of ${\dot M}_{acc}$ = 10 g s$^{-1}$ and values of ${\dot M}_{wind}$ of
 10$^{7}$, 10$^{6}$ and 10$^{5}$ g s$^{-1}$ shown by  the red, blue and green curves, respectively.  The downward pointing arrow represents the 
 upper bound to the mass of accumulated hydrogen in  GD 190 at the temperature range of the two 
 DBs in Table 1.  We see that in the plausible case that ${\dot M}_{wind}$ $<$ 10$^{6}$ g s$^{-1}$,  more
 hydrogen is predicted to be accumulated than observed.}
 \end{figure}

\begin{thebibliography}{}
	\bibitem [Alcock et al. (1986)] {Alcock1986} Alcock, C., Fristrom, C. C., \& Weigelman, R. 1986, \apj, 302, 462
	\bibitem [Asplund et al. (2004)] {Asplund2004} Asplund, M., Grevesse, N., Sauval, A. J., Allende Prieto, C., \& Kiselman, D. 2004, \aap, 417, 751
	\bibitem [Beauchamp et al. (1999)] {Beauchamp1999} Beauchamp, A., Wesemael, F., Bergeron, P., Fontaine, G., Saffer, R. A., Liebert, J., \& Brassard, P. 1999, \apj, 516, 887
	\bibitem [Bohlin et al. (1978)] {Bohlin1978} Bohlin, R. C., Savage, B. D., \& Drake, J. F. 1978, \apj, 224, 132
	\bibitem [Booth et al. (2009)] {Booth2009} Booth, M., Wyatt, M. C., Morbidelli, A., Moro-Martin, A., \& Levison, H. F. 2009, \mnras, 399, 385
\bibitem [Brinkworth et al. (2009)] {Brinkworth2009} Brinkworth, C. S., Gaensicke, B. T., Marsh, T. R., Hoard, D. W., \& Tappert, C. 2009, 696, 1402
\bibitem [Caffau et al. (2008)] {Caffau2008} Caffau, E., Ludwig, H.-G., Steffen, M., Ayres, T. R., Bonifacio, P., Cayrel, R., Freytag, B., \& Plez, B. 2008, \aap, 488, 1031
\bibitem [Cartledge et al. (2004)] {Cartledge2004} Cartledge, S. I. B., Lauroesch, J. T., Meyer, D. M., \& Sofia, U. J. 2004, \apj, 613, 1037
\bibitem [Delahaye \& Pinsonneault (2006)] {Delahaye2006} Delahaye, F., \& Pinsonneault, M. H. 2006, \apj, 649, 529
\bibitem [Desharnais et al. (2008)] {Desharnais2008} Desharnais, S., Wesemael, F., Chayer, P., Kruk, J. W., \& Saffer, R. A. 2008, \apj, 672, 540
\bibitem [Dones et al. (2004)] {Dones2004} Dones, L., Weissman, P. R., Levison,H. F., \& Duncan, M. J. 2004, in Comets II, ed. M. C. Festou, U. Keller, \& H. A. Weaver (Tucson, AZ: Univ. of Arizona), 153
\bibitem [Dufour et al. (2007)] {Dufour2007} Dufour, P., et al. 2007, \apj, 663, 1291
\bibitem [Dufour et al. (2010)] {Dufour2010} Dufour, P., Kilic, M., Fontaine, G., Bergeron, P., Lachapelle, F.-R., Kleinman, S. J., \& Leggett, S. K. 2010, \apj, 719, 803
\bibitem [Duncan et al. (1987)] {Duncan1987} Duncan, M., Quinn, T., \& Tremaine, S. 1987, \aj, 94, 1330
\bibitem [Dupuis et al. (1992)] {Dupuis1992} Dupuis, J., Fontaine, G., Pelletier, C., \& Wesemael, F. 1992, \apjs, 82, 505
\bibitem [Eisenstein et al. (2006)] {Eisenstein2006} Eisenstein, D. J. et al. 2006, \aj, 132, 676
\bibitem [Farihi et al. (2010)] {Farihi2010}  Farihi, J., Barstow, M., Redfield, S., Dufour, P., \& Hambly, N. C. 2010, \mnras, 404, 2123
\bibitem [Farihi et al. (2009)] {Farihi2009} Farihi, J., Jura, M., \& Zuckerman, B. 2009, \apj, 694, 805
\bibitem [Fontaine \& Brassard (2005)] {Fontaine2005} Fontaine, G., \& Brassard, P. 2005, in ASP Conf. Ser. 334, 14th European Workshop on White Dwarfs, ed. D. Koester \& S. Moehler (San Francisco, CA: ASP), 49
\bibitem [Francis (2005)] {Francis2005} Francis, P. J. 2005, \apj, 635, 1348
\bibitem [Gaensicke et al.  (2007)] {Gaensicke2007} Gaensicke, B. T., Marsh, T. R., \& Southworth, J. 2007, \mnras, 380, L35
\bibitem [Gaensicke et al. (2006)] {Gaensicke2006} Gaensicke, B. T., Marsh, T. R., Southworth, J., Rebassa-Mansergas, A. 2006, Science, 314, 1908
\bibitem [Gautschy \& Althaus (2002)] {Gautschy2002} Gautschy, A., \& Althaus, L. G. 2002, \aap, 382, 141
\bibitem [Gilmore \& Zeilik (2000)] {Gilmore2000} Gilmore, G., \& Zeilik, M. 2000, in Allen's Astrophysical Quantities, ed. A. N.Cox (New York: Springer), 471
\bibitem [Greenberg (1974)] {Greenberg1974} Greenberg, J. M. 1974, \apj, 189, L81
\bibitem [Greenberg (1998)] {Greenberg1998} Greenberg, J. M. 1998, \aap, 330, 375
\bibitem [Hansen \& Liebert (2003)] {Hansen2003} Hansen, B. M. S., \& Liebert, J. 2003, \araa, 41, 465
\bibitem [Jenkins (2009)] {Jenkins2009} Jenkins, E. B. 2009, \apj, 700, 1299
\bibitem [Jura (2003)]{Jura2003} Jura, M. 2003, \apj, 584, L91
\bibitem [Jura (2005)] {Jura2005} Jura, M. 2005, \aj, 130, 1261
\bibitem [Jura et al. (2009)] {Jura2009} Jura, M., Muno, M. P., Farihi, J., \& Zuckerman, B. 2009, \apj, 699, 1473
\bibitem [Jura \& Xu (2010)] {Jura2010} Jura, M. \& Xu, S. 2010, \aj, 140, 1129
\bibitem [Kilic et al. (2006)] {Kilic2006} Kilic, M., von Hippel, T., Leggett, S. K., \& Winget, D. E. 2006, \apj, 646, 474
\bibitem [Klein et al. (2010)] {Klein2010} Klein, B., Jura, M., Koester, D., Zuckerman, B., \& Melis, C. 2010, \apj, 709, 950
\bibitem [Koester (2009)] {Koester2009} Koester, D. 2009, \aap, 498, 517
\bibitem [Koester et al. (2005)] {Koester2005} Koester, D., Rollenhagen, K., Napiwotzki, R., Voss, B., Christlieb, N., Homeier, D., \& Reimers, D. 2005, \aap, 432, 1025
\bibitem [Kuijken \& Gilmore (1989)] {Kuijken1989} Kuijken, K., \& Gilmore, G. 1989, \mnras, 239, 571
\bibitem [Lamers \& Cassinelli (1999)] {Lamers1999} Lamers, H. H. G. L. M., \& Cassinelli, J. 1999, Introduction to Stellar Winds (Cambridge: Cambridge University Press)
\bibitem [Levison et al. (2010)] {Levison2010} Levison, H., Duncan, M. J., Brasser, R., \& Kaufmann, D. E. 2010, Science, 329, 187
\bibitem [Liebert et al. (1986)] {Liebert1986} Liebert, J., Wesemael, F., Hansen, C. J., Fontaine, G., Shipman, H L., Sion, E. M., Winget, D. E., \& Green, R. F. 1986, \apj, 309, 241
\bibitem [MacDonald \& Vennes (1991)] {Macdonald1991} MacDonald, J., \& Vennes, S. 1991, \apj, 371, 719
\bibitem [Marcy \& Butler (2000)] {Marcy2000} Marcy, G. W.,\& Butler, R. P. 2000, \pasp, 112, 137
\bibitem [McGlynn \& Chapman (1989)] {Mcglynn1989} McGlynn, T. A., \& Chapman, R. D. 1989, \apj, 346, L105
\bibitem [Melis et al. (2010)] {Melis2010} Melis, C., Jura, M., Albert, L., Klein, B., \& Zuckerman, B. 2010, \apj, 722, 1078
\bibitem [Musielak et al. (2003)] {Musielak2003} Musielak, Z. E., Noble, M., Porter, J. G., \& Winget, D. E. 2003, \apj, 593, 481
\bibitem [O'Dwyer et al. (2003)] {Odwyer2003} O'Dwyer, I. J., Chu, Y.-H., Gruendl, R. A., Guerrero, M. A., \& Webbink, R. F. 2003, \aj, 125, 2239
\bibitem [Parrtiott \& Alcock (1998)] {Parriott1998} Parriott, J., \& Alcock, C. 1998, \apj, 501, 357
\bibitem [Petitclerc et al. (2005)] {Petitclerc2005} Petitclerc, N., Wesemael, F., Kruk, J. W., Chayer, P., \& Billeres, M. 2005, \apj, 624, 317
\bibitem [Plavchan et al. (2005)] {Plavchan2005} Plavchan, P., Jura, M., \& Lipscy, S. 2005, ApJ, 631, 1161
\bibitem [Provencal et al. (1996)] {Provencal1996} Provencal, J. L., Shipman, H. L., Thejll, P., Vennes, S., \& Bradley, P. A. 1996, \apj, 466, 1011
\bibitem [Provencal et al. (2000)] {Provencal2000} Provencal, J. L., Shipman, H. L., Thejll, P., \& Vennes, S. 2000, \apj, 542, 1041
\bibitem [Provencal et al. (2009)] {Provencal2009} Provencal, J. L. et al. 2009, \apj, 693, 564
\bibitem [Reach et al. (2010)] {Reach2010} Reach, W. T., Vaubaillon, J., Lisse, C. M., Holloway, M., \& Rho, J. 2010, Icarus, 208, 276
\bibitem [Sekanina (1976)] {Sekanina1976} Sekanina, Z. 1976, Icarus, 27, 123
\bibitem [Sekanina (1982)] {Sekanina1982} Sekanina, Z. 1982, in Comets, ed. L. Wilkening (Tucson: University of Arizona), 251
\bibitem [Sekanina (2002)] {Sekanina2002} Sekanina, Z. 2002, \apj, 566, 577
\bibitem [Sen \& Rama (1993)] {Sen1993} Sen, A. K., \& Rama, N. C. 1993, \aap, 275, 298
\bibitem [Shull \& Stern (1995)] {Shull1995} Shull, J. M., \& Stern, S. A. 1995, \aj, 109, 690
\bibitem [Siebert et al. (2003)] {Siebert2003} Siebert, A., Bienayme, O., \& Soubiran, C. 2003, \aap, 399, 531
\bibitem [Sion et al. (1988)] {Sion1988} Sion, E. M., Fritz, M. L., McMullin, J. P., \& Lallo, M. D. 1988, \aj, 96, 251
\bibitem [Spitzer (1978)] {Spitzer1978} Spitzer, L. 1978, Physical Processes in the Interstellar Medium (New York: J. Wiley)
\bibitem [Stern (1990)] {Stern1990a} Stern, S. A. 1990, \pasp, 102, 793
\bibitem [Stern \& Shull (1990)] {Stern1990b} Stern, S. A., \& Shull, J. M. 1990, \apj, 359, 506
\bibitem [Su \& Li (2010)] {Su2010} Su, J., \& Li, Y. 2010, Res. Astron. Astrophys. 10, 266
\bibitem [Sullivan et al. (2008)] {Sullivan2008} Sullivan, D. J. et al. 2008, \mnras, 387, 137
\bibitem [Tremblay et al. (2008)] {Tremblay2008} Tremblay, P.-E., \& Bergeron, P. 2008, \apj, 672, 1144
\bibitem [Unglaub (2008)] {Unglaub2008} Unglaub, K. 2008, \aap, 486, 923
\bibitem [Unglaub \& Bues (2000)] {Unglaub2000} Unglaub, K., \& Bues, I. 2000, \aap, 359, 1042
\bibitem [Vennes et al. (2010)] {Vennes2010} Vennes, S., Kawka, A., \& Nemeth, P. 2010, \mnras, 404, L40
\bibitem [von Hippel et al. (2007)] {Vonhippel2007} von Hippel, T., Kuchner, M. J., Kilic, M., Mullally, F., \& Reach, W. T. 2007, \apj, 662, 544
\bibitem [Voss et al. (2007)] {Voss2007} Voss, B., Koester, D., Napiwotzki, R., Christlieb, N., \& Reimers, D. 2007, \aap, 470, 1079
\bibitem [Weissman (1996)] {Weissman1996}Weissman, P. R. 1996, in ASP Conf. Ser. 107, Completing the Inventory of the Solar System, ed. T. W. Rettig \& J. M. Hahn (San Francisco: ASP) 265
\bibitem [Whipple (1975)] {Whipple1975} Whipple, F. L. 1975, \aj, 80, 525
\bibitem [Whittet (2010)] {Whittet2010} Whittet, D. C. B. 2010, \apj, 710, 1009
\bibitem [Zuckerman et al. (2007)] {Zuckerman2007} Zuckerman, B., Koester, D., Melis, C., Hansen, B., \& Jura, M. 2007, \apj, 671, 872
\bibitem [Zuckerman et al. (2010)] {Zuckerman2010} Zuckerman, B., Melis, C., Klein, B., Koester, D., \& Jura, M. 2010, \apj, 722, 725
\end{thebibliography}
\end{document}